\documentclass[a4paper,12pt]{article}
\usepackage{graphicx,amssymb,bm}

\makeatletter
\@addtoreset{equation}{section}

\makeatother
\newcommand{\bea}{\begin{eqnarray}}
\newcommand{\ena}{\end{eqnarray}}
\newcommand{\vs}[1]{\vspace{#1 mm}}
\newcommand{\hs}[1]{\hspace{#1 mm}}
\renewcommand{\a}{\alpha}
\renewcommand{\b}{\beta}
\renewcommand{\c}{\gamma}
\renewcommand{\d}{\delta}
\newcommand{\e}{\epsilon}
\newcommand{\s}{\sigma}

\def\bbox{{\,\lower0.9pt\vbox{\hrule \hbox{\vrule height 0.2 cm
\hskip 0.2 cm \vrule height 0.2 cm}\hrule}\,}}
\newcommand{\dsl}{\pa \kern-0.5em /}

\newcommand{\shalf}{\frac{1}{2}}
\newcommand{\pa}{\partial}

\newcommand{\nn}{\nonumber\\}
\newcommand{\p}[1]{(\ref{#1})}

\begin{document}

\renewcommand{\thefootnote}{\fnsymbol{footnote}}
\begin{titlepage}

\setcounter{page}{0}
\begin{flushright}
OU-HET 438 \\
hep-th/0304172
\end{flushright}

\vs{10}
\begin{center}
{\Large\bf A Study of Accelerating Cosmologies from Superstring/M Theories}
\vs{15}

{\large
Nobuyoshi Ohta\footnote{e-mail address: ohta@phys.sci.osaka-u.ac.jp}} \\
\vs{10}
{\em Department of Physics, Osaka University,
Toyonaka, Osaka 560-0043, Japan}

\end{center}
\vs{15}
\centerline{{\bf{Abstract}}}
\vs{5}

We study aspects of the accelerating cosmologies obtained from
the compactification of vacuum solution and S2-branes of superstring/M
theories. Parameter dependence of the resulting expansion of our universe
and internal space is examined for all cases.
We find that accelerated expansions are obtained also from spherical
internal spaces, albeit the solution enters into contracting phase eventually.
The relation between the models of SM2- and SD2-branes are also discussed,
and a potential problem with SD2-brane model is noted.

\end{titlepage}
\newpage
\renewcommand{\thefootnote}{\arabic{footnote}}
\setcounter{footnote}{0}
\setcounter{page}{2}

\section{Introduction}

Recent discovery of the cosmic acceleration calls for explanation within
the framework of fundamental theory of superstring/M theories. It has been
known for some time that it is difficult to derive such cosmologies from
the compactifications of solutions in superstring/M theories
and has been considered that there is a no-go theorem which excludes such
possibility if one chooses the internal space compact and static
manifolds~\cite{G}.

Recently it has been shown that this no-go theorem can be overcome
if one allows time-dependence, and a solution of the
vacuum Einstein equations with compact hyperbolic internal space
has been proposed for such a model~\cite{TW}.
The solution turned out~\cite{NO1} to be a special case of the time-dependent
solutions discovered before~\cite{NO2}. Moreover it has been
found~\cite{NO1} that similar accelerating cosmologies can be obtained
for S(pacelike)M2-branes~\cite{GS,CGG,KMP,DK,NO2} and that not only the
hyperbolic but also flat internal spaces give similar behavior.
The SM2-brane case is also studied in ref.~\cite{WO}. Later SD2-brane,
which can be obtained from SM2-brane by dimensional reduction, has been
discussed in ref.~\cite{R}. However, problem with strong string coupling
in SD2-brane casts a question on the validity of the results at least for
flat internal space. Related cosmologies with accelerated expansion are
also discussed in refs.~\cite{M,CC}.

In ref.~\cite{WO}, the degree of the expansion factor during the period of
the accelerated expansion has been studied for SM2-brane, with the small
result of order 2. This appears to be too small to solve cosmological
issues such as the horizon or flatness problems, which require the magnitude
${\cal O}(e^{60})$. Since the solutions contain some parameters, it is
interesting to examine if there is a possibility of evading this problem.

In this paper, we study aspects of these models of accelerated
expansions including parameters for vacuum, SM2-brane and SD2-brane solutions.
We also estimate how the expansion factors change by varying the parameters.
It turns out that the factors can change but not enough to get around
the difficulty mentioned above. We also find that it is possible to have
accelerated expansion for spherical internal space too for S2-branes
(but not for vacuum solutions) if one chooses parameters appropriately,
although the universe enters into contracting phase after some period.
We further discuss how SD2-branes are obtained from SM2-branes by
dimensional reduction. One can thus see that the features of the
cosmologies from SD2-brane is basically similar to those from SM2-brane.
In fact we find that qualitative behavior does not change much, as expected.
However, there is a potential problem with the SD2-brane
model because it often suffers from the strong string coupling and
then it is not clear if we can trust the result in this picture.
It is known that the strong string coupling can be understood as the
11-th dimension becoming large, so that the description by M theory
becomes more relevant~\cite{WI}.

This paper is organized as follows. In sect.~2, we first discuss the
relation of the vacuum and our S-brane solutions, and study the cosmologies
from the vacuum solutions. In this case, there is no adjustable parameter
and the result does not change.
In sect.~3, we examine the cosmologies from SM2-brane. The hyperbolic, flat
and spherical internal spaces are discussed in subsections~3.1, 3.2 and 3.3,
respectively. We study these cases by changing a parameter, and find
that an interesting model can be obtained also from spherical internal space.
In sect.~4, the way how SD2-brane is obtained from SM2-brane is explicitly
shown. We discuss the cosmologies in this case briefly.
We also note that in many cases in this kind of models, the string coupling
becomes strong so that the picture becomes obscure. We show how this
can be avoided for compactification on hyperbolic manifolds, but there
seems to be no way to avoid this for flat and spherical manifolds.
Finally sect.~5 is devoted to conclusions and discussions.

In the course of writing this article, a paper appeared with related
observations on the spherical internal space for SM2-brane~\cite{EG}.

\section{Vacuum solution and S2-brane}

\subsection{Generality}

It is convenient to write the $(4+n)$-dimensional solution as
\bea
ds^2 = \d^{-n}(t) ds_E^2 + \d^2(t) d\Sigma_{n,\s}^2,
\label{dsol}
\ena
where $n$ is the dimension of the internal spherical ($\s=+1$),
flat ($\s=0$) or hyperbolic ($\s=-1$) spaces, whose line elements are
$d\Sigma_{n,\s}^2$, and
\bea
ds_E^2 = -S^6(t) dt^2 + S^2(t) d{\bf x}^2,
\label{4sol}
\ena
describes the 4-dimensional spacetime. The form of \p{dsol} is chosen
such that the metric in \p{4sol} are in the Einstein frame. The solution
in refs.~\cite{TW} is given as
\bea
\d(t) &=& e^{-3t/(n-1)} \left(\frac{\sqrt{3(n+2)/n}}{(n-1)
\sinh(\sqrt{3(n+2)/n}\; |t|)}\right)^{\frac{1}{(n-1)}}, \nn
S(t) &=& e^{-(n+2)t/2(n-1)} \left(\frac{\sqrt{3(n+2)/n}}{(n-1)
\sinh(\sqrt{3(n+2)/n}\; |t|)}\right)^{\frac{n}{2(n-1)}}.
\label{vfactor}
\ena
with hyperbolic internal space.

If we take the time coordinate $\eta$ defined by
\bea
d\eta= S^3(t) dt,
\label{time}
\ena
the metric~\p{4sol} describes a flat homogeneous isotropic universe with
scale factor $S(t)$, and $\d(t)$ gives the measure of the size of internal
space. The condition for expanding 4-dimensional universe is that
\bea
\frac{dS}{d\eta}>0.
\label{cond1}
\ena
Accelerated expansion is obtained if, in addition,
\bea
\frac{d^2S}{d\eta^2}>0.
\label{cond2}
\ena
It has been shown that these can be satisfied for $n=7$ and for certain
period of negative $t$ (with the convention $t_1=0$) which is the period
that our universe is evolving ($t<0$ and $t>0$ are two disjoint possible
universes)~\cite{TW}.

We have noted~\cite{NO1} that the above solution is actually a special
case of the solutions derived in ref.~\cite{NO2}. The theory considered
there is given by the action for $d$-dimensional gravity coupled to a
dilaton $\phi$ and $m$ different $n_A$-form field strengths:
\bea
I = \frac{1}{16 \pi G_d} \int d^d x \sqrt{-g} \left[
R - \shalf (\pa \phi)^2 - \sum_{A=1}^m \frac{1}{2 n_A!} e^{a_A \phi}
F_{n_A}^2 \right].
\label{act}
\ena
This action describes the bosonic part of $d=11$ or $d=10$ supergravities;
we simply drop $\phi$ and put $a_A=0$ and $n_A=4$ for $d=11$, whereas we
set $a_A=-1$ for the NS-NS 3-form and $a_A=\shalf(5-n_A)$ for forms coming
from the R-R sector.
The field strength for an electrically charged S$q$-brane is given by
\bea
F_{t \a_1 \cdots \a_{q+1}} = \e_{\a_1 \cdots \a_{q+1}} \dot E, \hs{3}
(n_A = q+2),
\label{ele}
\ena
where $\a_1, \cdots ,\a_{q+1}$ stand for the tangential directions to
the S$q$-brane. The magnetic case is given by
\bea
F^{\a_{q+2} \cdots \a_p a_1 \cdots a_{n}} = \frac{1}{\sqrt{-g}}
e^{-a\phi} \e^{t \a_{q+2} \cdots \a_p a_1 \cdots a_n} \dot E,
\hs{3} (n_A = d-q-2)
\label{mag}
\ena
where $a_1, \cdots, a_{n}$ denote the coordinates of the $n$-dimensional
hypersurface $\Sigma_{n,\s}$.

In ref.~\cite{CGG} a single S-brane solution was given, and in
ref.~\cite{NO2} general orthogonally intersecting solutions were derived
by solving field equations. Our solutions restricted to a single S$q$-brane
with $(q+1)$-dimensional world-volume in $p$-dimensional space~\cite{NO2} are
(hereafter the subscript $A$ is not necessary and is dropped)
\bea
\label{oursol}
ds_d^2 &=& [\cosh\tilde c (t-t_2)]^{2 \frac{q+1}{\Delta}}
\Bigg[ e^{2c_0 t+2c_0'} \left\{ - e^{2ng(t)} dt^2
+ e^{2g(t)}d\Sigma_{n,\s}^2\right\} \nn
&& \hs{20} +\; \sum_{\a=1}^p [\cosh\tilde c(t-t_2)]^{- 2
\frac{\c^{(\a)}}{\Delta}} e^{2 \tilde c_\a t+2c_\a'} (dx^\a)^2\Bigg], \\
E &=& \sqrt{\frac{2(d-2)}{\Delta}}\frac{e^{\tilde c(t-t_2)
-\e a c_\phi'/2+\sum_{\a\in q} c_\a'}}{\cosh \tilde c(t-t_2)},\quad
\tilde c = \sum_{\a\in q} c_\a-\frac{1}{2} c_\phi \e a, \nn
\phi &=& \frac{(d-2)\e a}{\Delta} \ln \cosh\tilde c(t-t_2)
+\tilde c_\phi t + c_\phi',
\label{oursol1}
\ena
where $d=p+n+1$ and $\e= +1 (-1)$ corresponds to electric (magnetic) fields.
The coordinates $x^\a, (\a=1,\ldots, p)$ parametrize the $p$-dimensional
space, within which $(q+1)$-dimensional world-volume of S$q$-brane is
embedded, and the remaining coordinates of the $d$-dimensional spacetime
are the time $t$ and coordinates on compact $n$-dimensional spherical
($\s=+1$), flat ($\s=0$) or hyperbolic ($\s=-1$) spaces.
We have also defined
\bea
\label{res2}
&& \Delta = (q + 1) (d-q-3) + \shalf a^2 (d-2), \nn
&& \c^{(\a)} = \left\{ \begin{array}{l}
d-2 \\
0
\end{array}
\right.
\hs{5}
{\rm for} \hs{3}
\left\{
\begin{array}{l}
x_\a \hs{3} {\rm belonging \hs{2} to} \hs{2} q{\rm -brane} \\
{\rm otherwise}
\end{array},
\right.
\ena
and
\bea
g(t) = \left\{\begin{array}{ll}
\frac{1}{n-1} \ln \frac{\b}{\cosh[(n-1)\b(t-t_1)]} & :\s=+1, \\
\pm \b(t-t_1) & :\s=0, \\
\frac{1}{n-1} \ln \frac{\b}{\sinh[(n-1)\b|t-t_1|]} & :\s=-1,
\end{array}
\right.
\label{ints}
\ena
$\b, t_1, t_2$ and $c$'s are integration constants which satisfy
\bea
&& c_0 = \frac{q+1}{\Delta}\tilde c -\frac{\sum_{\a=1}^{p} c_\a}{n-1}, \;\;
c_0' = -\frac{\sum_{\a=1}^{p} c_\a'}{n-1}, \nn
&& \tilde c_\a = c_\a - \frac{\c^{(\a)}-q-1}{\Delta}\tilde c, \;\;
\tilde c_\phi = c_\phi + \frac{(d-2)\e a}{\Delta}\tilde c.
\label{intconst1}
\ena
These must further obey the condition
\bea
\frac{1}{n-1}\left(\sum_{\a=1}^{p} c_\a\right)^2
+ \sum_{\a=1}^{p} c_\a^2 + \shalf c_\phi^2= n(n-1) \b^2.
\label{condconst}
\ena
The free parameters in our solutions are $c_\a, c_\a' (\a=1,\cdots,p),
c_\phi, c_\phi',t_1$ and $t_2$.
The time derivative of $E$ gives the field strengths of antisymmetric
tensor and in our convention they are given as
\bea
\left. \begin{array}{c}
e^{a\phi}*\! F \\
F
\end{array}
\right\}
=\tilde c \;\sqrt{\frac{2(d-2)}{\Delta}}
e^{-\sum_{\a\in q}c'_\a+\e ac_\phi'/2} dx^{\a_{q+2}}\wedge \cdots \wedge
dx^{\a_p}\wedge \;\mbox{Vol}(\Sigma_{n,\s}),
\ena
for electric (first line) and magnetic (second line) fields, where
Vol$(\Sigma_{n,\s})$ is the unit volume form of the hypersurface
$\Sigma_{n,\s}$ and $*$ represents dual.
We can check that $\sqrt{\frac{2(d-2)}{\Delta}}=1$ for SM- and SD-branes.
The above solution includes that in ref.~\cite{CGG}, and the precise
relation is given in \cite{NO2}. (If we also use the relation
$c_\phi'=\frac{d-2}{q-1} ac_\c'+c_2, (\c=1, \cdots, p)$ which was not given
in eq.~(40) there, we reproduce $F=b\; dy^1 \wedge \cdots \wedge
dy^{q-n}\wedge \mbox{Vol}(\Sigma_{n,\s})$.)

For the general S2-brane obtained from the solution~\p{oursol} by putting
$p=q+1=3, c\equiv c_1=c_2=c_3, c'\equiv c_1'=c_2'=c_3'$, we find that
it takes the form~\p{dsol} and \p{4sol} with
\bea
\d(t) &=& [\cosh \tilde c(t-t_2)]^{3/\Delta} e^{g(t) + c_0 t + c_0'}, \nn
S(t) &=& [\cosh \tilde c(t-t_2)]^{(n+2)/2\Delta} e^{ng(t)/2 +(n+2)(c_0 t
+ c_0')/6},
\label{genfactor}
\ena
where
\bea
&& \tilde c = 3c-\frac{1}{2} c_\phi \e a, \quad
c_0 = \frac{3}{\Delta}\tilde c - \frac{3}{n-1}c, \quad
c_0' =-\frac{3}{n-1}c', \nn
&& \b = \sqrt{\frac{3(n+2)}{n(n-1)^2} c^2 + \frac{1}{2n(n-1)}c_\phi^2}.
\ena
We are now going to discuss how this S2-brane solution includes the vacuum
solution~\p{vfactor} and the resulting cosmologies~\cite{TW}.

\subsection{Vacuum solution}

The relation between $\tilde c$ and $c_\a$ and $c_\phi$ in eq.~\p{oursol1}
is derived~\cite{NO2} under the assumption that we have the independent field
strengths $F$. In the absence of these, we can disregard this relation
and should set $\tilde c$ to zero.
It is then easy to see that the solution~\p{oursol} reproduces
\p{dsol}-\p{vfactor} for $p=q+1=3, \s=-1, c=1, c'=0$ without dilaton
($c_\phi=0$)~\cite{NO1}. The scale factor is simply~\p{genfactor} with
$\tilde c=0$ which coincides with \p{vfactor}. On the other hand,
the S-brane solutions derived in \cite{CGG} assume nonzero field strengths
from the start, and one cannot simply get the vacuum solution.

The condition of the expansion~\p{cond1} for the vacuum solution~\p{vfactor}
is~\cite{TW}
\bea
n_1(t) \equiv -1-\sqrt{\frac{3n}{n+2}}\coth\left(\sqrt{\frac{3(n+2)}{n}}\;
c(t-t_1)\right)>0,
\label{cond01}
\ena
where we have also included parameters $c,c'$ and $t_1$.
The condition~\p{cond2} gives
\bea
\frac{3(n-1)}{(n+2)\sinh^2[\sqrt{3(n+2)/n}\; c(t-t_1)]} - n_1^2(t) > 0.
\label{cond02}
\ena
The parameter $t_1$ and $c$ can be absorbed into the shift and rescaling
of the time $t$. Hence without loss of generality, we can set $t_1=0$ and
$c=1$ (changing $c$ gives the change in the scale of time).
There is a singularity in $S(t)$ at $t=0$, but the time $\eta$ run from 0
to infinity while $t$ runs from $-\infty$ to 0, which is an infinite
future for any event with $t<0$ and hence the evolution of our universe
can be restricted to $t<0$.

\begin{figure}[htb]
\begin{minipage}{.45\linewidth}
\begin{center}
\setlength{\unitlength}{.7mm}
\begin{picture}(50,45)(20,5)
\includegraphics[width=6cm]{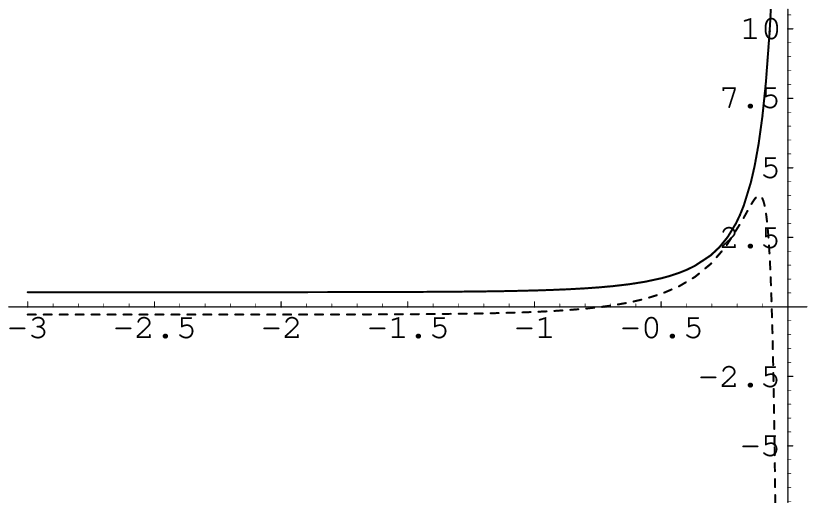}
\put(5,20){\footnotesize $t$}
\end{picture}
\caption{The lhs of eq.~\p{cond01} (solid line) and \p{cond02} (dashed line).}
\label{f1}
\end{center}
\end{minipage}
\hspace{3mm}
\begin{minipage}{.5\linewidth}
\begin{center}
\setlength{\unitlength}{.7mm}
\begin{picture}(50,45)(15,5)
\includegraphics[width=6cm]{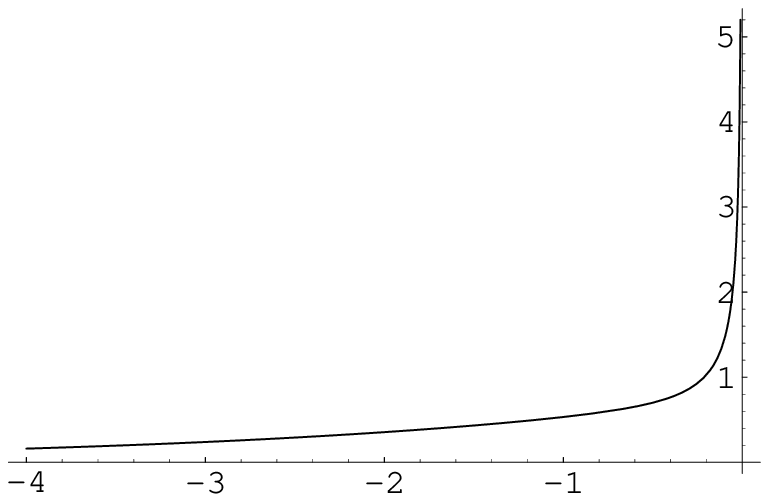}
\put(0,47){\footnotesize $S(t)$}
\put(2,4){\footnotesize $t$}
\end{picture}
\caption{The behavior of the scale factor $S(t)$ in~\p{vfactor}.}
\label{f2}
\end{center}
\end{minipage}
\end{figure}

The left hand side of these eqs. for $n=7$ are shown in Fig.~\ref{f1},
and the behavior of the scale factor $S(t)$ in~\p{vfactor} is depicted in
Fig.~\ref{f2}.
We see that there is a certain period of negative time
that the conditions~\p{cond01} and \p{cond02} are satisfied~\cite{TW}.
The period of the accelerated expansion can be adjusted by changing the
constant $c$, but this does not affect the resulting expansion factor.
The scale factor vanishes in the infinite past, but diverges in the infinite
future.

The expansion factor $A$ during the accelerated expansion is given by the
ratio of $S(t)$ at the starting time $T_1$ and ending time $T_2$ of the
acceleration. These are read off from Fig.~\ref{f1} as
\bea
T_1 \simeq -0.73, \qquad
T_2 \simeq -0.07.
\ena
If we keep the parameter $c$, $T=ct$ and this simply changes the scale of
the time by a constant multiplicative factor. We then find that the expansion
factor is
\bea
A =\frac{S(T_2)}{S(T_1)} \simeq 2.91.
\ena
This value is too small to explain the cosmological problems.
Note that there is no parameter to improve $A$ here.

\begin{figure}[htb]
\begin{center}
\setlength{\unitlength}{.7mm}
\begin{picture}(50,45)(20,5)
\includegraphics[width=6cm]{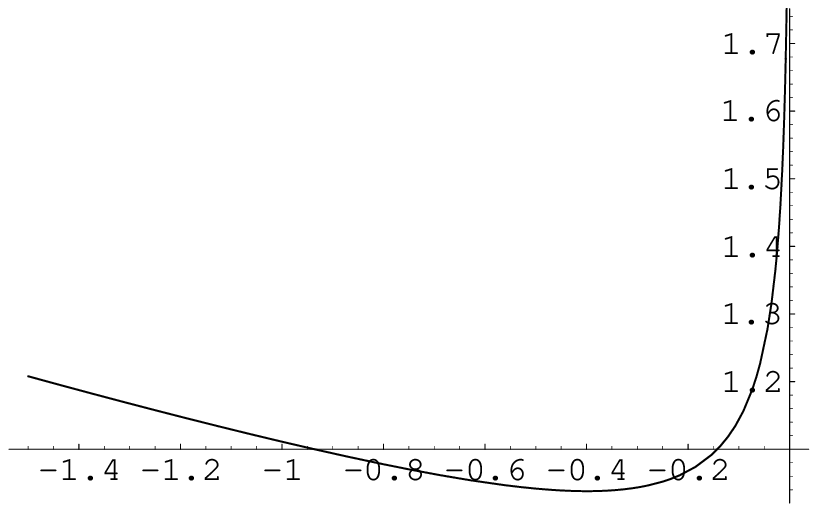}
\put(2,47){\footnotesize $\d(t)$}
\put(5,5){\footnotesize $t$}
\end{picture}
\caption{The behavior of the size of the internal space $\d(t)$
in~\p{vfactor}.}
\label{f3}
\end{center}
\end{figure}

The behavior of the size of the internal space is also shown in Fig.~\ref{f3}.
Though it appears as if the internal space rapidly expands in terms of
$t$, the actual time coordinate is $\eta$ and the region $t\sim 0^-$ is
actually long.
When the acceleration starts, $\d(t)$ shrinks but eventually starts expanding,
and the ratio of the sizes during the accelerated expansion is 2.18.
As observed for SM2-brane case~\cite{WO}, there is no stable point in the
size of the internal space, and in the infinite future and past its
size goes to infinity. This seems to be the common problem in the hyperbolic
compactification~\cite{NSS}. We will find this behavior in other cases.

We have also examined other internal spaces, but without any adjustable
parameter we find that neither flat nor spherical spaces give accelerating
cosmologies~\cite{NO1}.

\section{SM2-brane}

The SM2-brane in M-theory can be obtained from \p{oursol} by putting
$p=q+1=3, \e=+1, a=0, c_\phi=0$ without dilaton and $\Delta =3(n-1)$.
We also put $c\equiv c_1=c_2=c_3, c'\equiv c_1'=c_2'=c_3'$ and then
$\b=\frac{1}{n-1}\sqrt{\frac{3(n+2)}{n}}\; c$ is determined from \p{condconst}.
The solution~\p{oursol} then gives
\bea
ds_d^2 &=& [\cosh 3c(t-t_2)]^{2/(n-1)} \Bigg[ -e^{2ng(t)-6c'/(n-1)} dt^2
+ e^{2g(t)-6c'/(n-1)}d\Sigma_{n,\s}^2 \nn
&& \hs{20} +\; [\cosh 3c(t-t_2)]^{- 2(n+2)/3(n-1)2} e^{2c'} d{\bf x}^2\Bigg].
\ena
This is the universe~\p{dsol}-\p{4sol} with
\bea
\d(t) &=& [\cosh3c(t-t_2)]^{1/(n-1)} e^{g(t) -3c'/(n-1)}, \nn
S(t) &=& [\cosh3c(t-t_2)]^{(n+2)/6(n-1)} e^{ng(t)/2 -(n+2)c'/2(n-1)}.
\label{mfactor}
\ena
We now discuss three internal spaces~\p{ints} separately.

\subsection{Hyperbolic internal space}

The conditions~\p{cond1} and \p{cond2} for $n=7$ of our interest and
hyperbolic internal space $\s=-1$ are (again shifting the time to set $t_1=0$)
\bea
\label{cond11}
&& n_2(t) \equiv \frac{3}{4} \tanh[3c(t-t_2)]-\frac{\sqrt{21}}{4}
\coth(3\sqrt{3/7}ct)>0, \\
&& \frac{9}{8}\left(\frac{1}{\cosh^2[3c(t-t_2)]}
+ \frac{1}{\sinh^2(3\sqrt{3/7}ct)}\right) - n_2^2(t) > 0.
\label{cond12}
\ena

\begin{figure}[htb]
\begin{minipage}{.45\linewidth}
\begin{center}
\setlength{\unitlength}{.7mm}
\begin{picture}(50,45)(20,5)
\includegraphics[width=6cm]{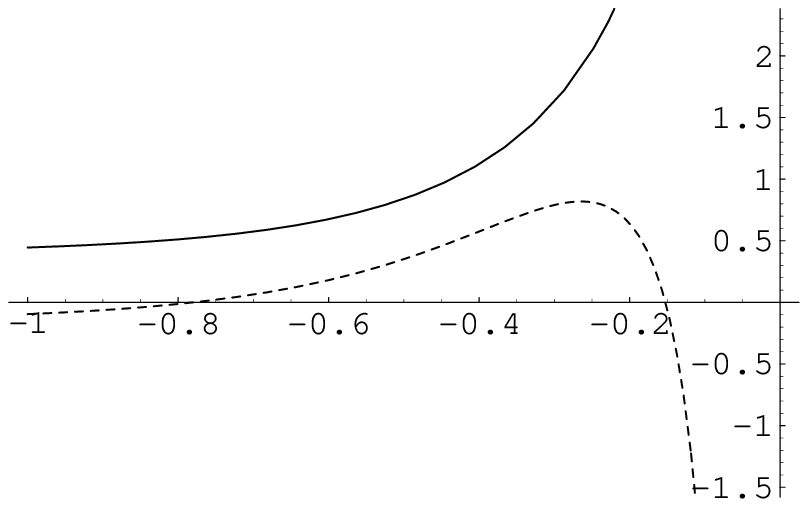}
\put(5,20){\footnotesize $t$}
\end{picture}
\caption{The lhs of eq.~\p{cond11} (solid line) and \p{cond12} (dashed line).}
\label{f4}
\end{center}
\end{minipage}
\hspace{3mm}
\begin{minipage}{.5\linewidth}
\begin{center}
\setlength{\unitlength}{.7mm}
\begin{picture}(50,45)(15,5)
\includegraphics[width=6cm]{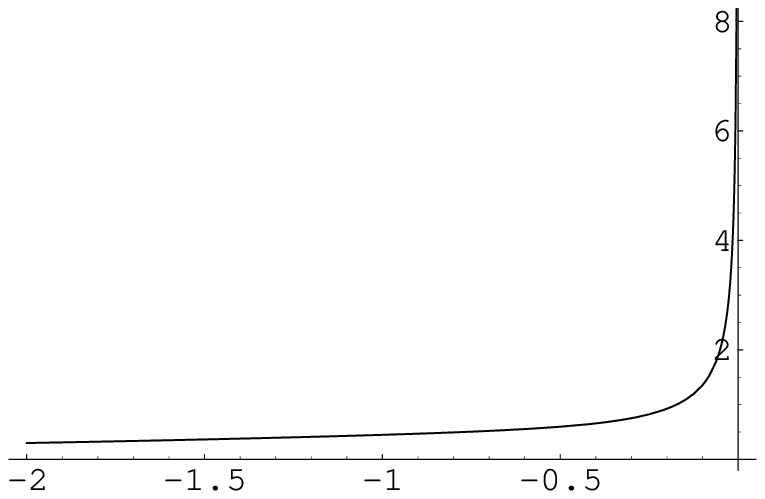}
\put(0,50){\footnotesize $S(t)$}
\put(2,2){\footnotesize $t$}
\end{picture}
\caption{The behavior of the scale factor $S(t)$ in~\p{mfactor}.}
\label{f5}
\end{center}
\end{minipage}
\end{figure}

Here we can again consider that our universe evolves only for $t<0$.
The left hand side of these eqs. for $c=1$ and $t_2=0$ are shown in
Fig.~\ref{f4}, and the behavior of the scale factor $S(t)$ in~\p{mfactor} is
depicted in Fig.~\ref{f5}. We again see that there is a certain period of
negative time that these conditions are satisfied~\cite{NO1}.

The starting time $T_1$ and ending time $T_2$ of the acceleration are
read off from Fig.~\ref{f4} as~\cite{WO}
\bea
T_1 \simeq -0.78, \qquad
T_2 \simeq -0.15.
\ena
We then find that the expansion factor is
\bea
A \simeq 2.17.
\ena
This value is too small to explain the cosmological problems.
However, here is a parameter $t_2$ in contrast to the vacuum solution,
and it is interesting to check what effect this may have.

We have examined the conditions for various choices of $t_2$ and found that
the typical behavior for positive $t_2$ is basically the same as $t_2=0$
case, but the period of the accelerated expansion changes slightly.
For example, for $t_2=1$, we find
\bea
T_1 \simeq -0.73, \qquad
T_2 \simeq -0.06.
\ena
and the value of the expansion factor during the accelerated expansion
improves:
\bea
A \simeq 3.13.
\ena
This is still not enough improvement for cosmological applications.
Increasing the value of $t_2$ does not affect the numerical value of $A$
much. The behavior for negative $t_2$ is also basically the same, but the
expansion factor is worse; one typically gets $A \sim 1.4$.

\begin{figure}[htb]
\begin{center}
\setlength{\unitlength}{.7mm}
\begin{picture}(50,45)(20,5)
\includegraphics[width=6cm]{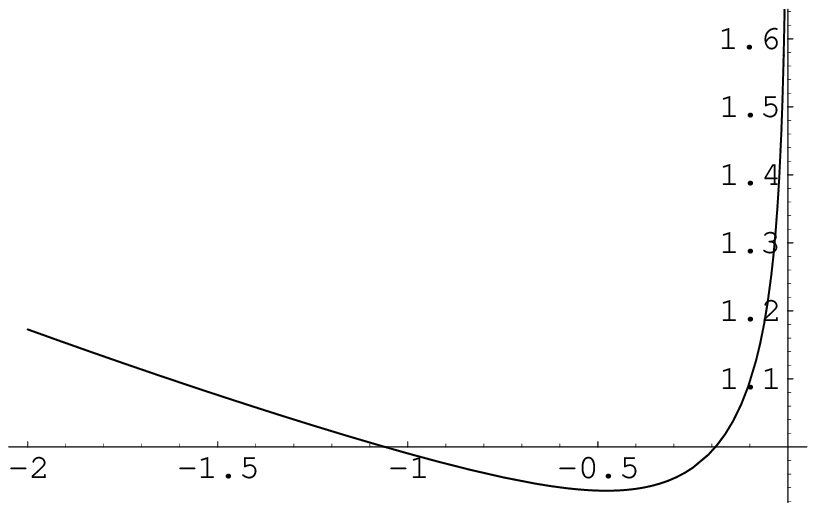}
\put(2,50){\footnotesize $\d(t)$}
\put(5,5){\footnotesize $t$}
\end{picture}
\caption{The behavior of the size of the hyperbolic internal space $\d(t)$
in~\p{mfactor}.}
\label{f6}
\end{center}
\end{figure}

The behavior of the size of the internal space for $t_2=0$ is also shown
in Fig.~\ref{f6}.
As observed in~\cite{WO}, there is no stable point in the size of the
internal space, and in the infinite future its size goes to infinity.
There is no significant change in this behavior if we change the parameter
$t_2$.

\subsection{Flat internal space}

The conditions~\p{cond1} and \p{cond2} for $n=7$ and flat internal
space $\s=0$ are
\bea
\label{cond21}
&& n_3(t) \equiv \frac{3}{4} \tanh[3c(t-t_2)]+\frac{\sqrt{21}}{4}>0, \\
&& \frac{9}{8}\frac{1}{\cosh^2[3c(t-t_2)]} - n_3^2(t) > 0,
\label{cond22}
\ena
where we have chosen the plus sign in eq.~\p{ints} since minus sign cannot
give expanding universe.

\begin{figure}[htb]
\begin{minipage}{.45\linewidth}
\begin{center}
\setlength{\unitlength}{.7mm}
\begin{picture}(50,45)(20,5)
\includegraphics[width=6cm]{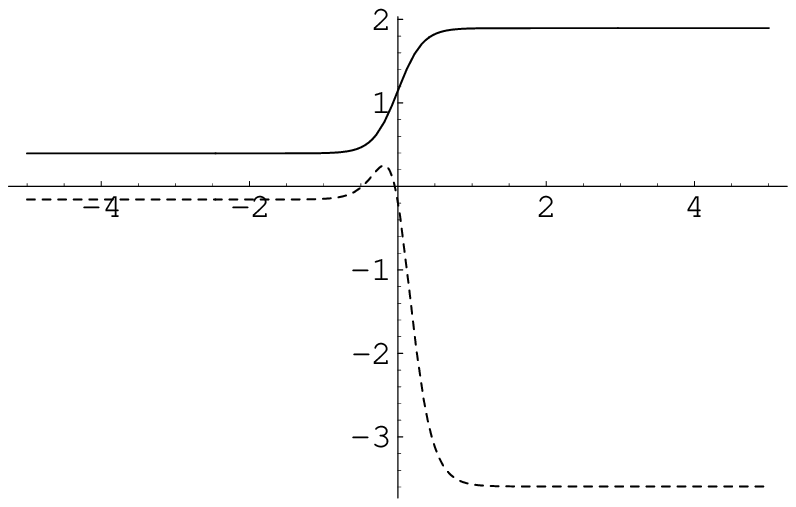}
\put(5,30){\footnotesize $t$}
\end{picture}
\caption{The lhs of eq.~\p{cond21} (solid line) and \p{cond22} (dashed line).}
\label{f7}
\end{center}
\end{minipage}
\hspace{3mm}
\begin{minipage}{.5\linewidth}
\begin{center}
\setlength{\unitlength}{.7mm}
\begin{picture}(50,45)(15,5)
\includegraphics[width=6cm]{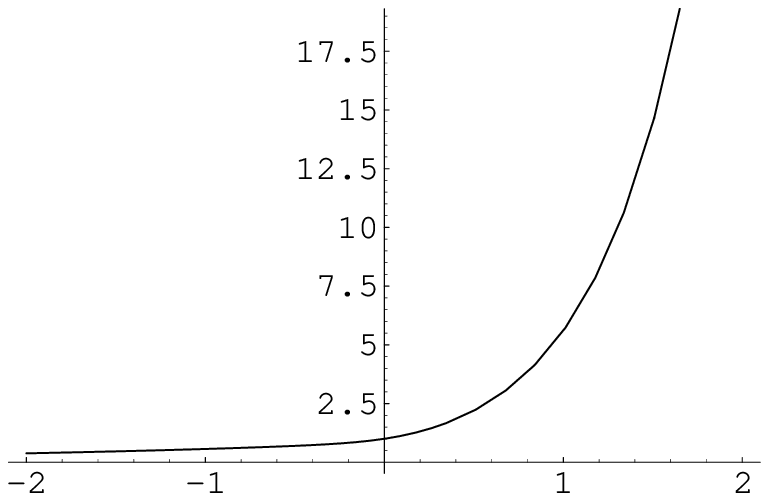}
\put(-40,50){\footnotesize $S(t)$}
\put(3,2){\footnotesize $t$}
\end{picture}
\caption{The behavior of the scale factor $S(t)$ in~\p{mfactor}.}
\label{f8}
\end{center}
\end{minipage}
\end{figure}

Here since $S(t)$ does not have any singularity and is positive, the time
$t$ runs from $-\infty$ to $+\infty$ while the time $\eta$ runs from 0
to $+\infty$ monotonously. The left hand side of these eqs. for $c=1$ and
$t_2=0$ are shown in Fig.~\ref{f7}, and the behavior the scale factor
$S(t)$ in~\p{mfactor} is depicted in Fig.~\ref{f8}. We again see that there
is a certain period of negative time that these conditions are
satisfied~\cite{NO1}.

We again read off the starting time $T_1$ and ending time $T_2$ of the
acceleration from Fig.~\ref{f7} as
\bea
T_1 \simeq -0.48, \qquad
T_2 \simeq -0.04,
\ena
and the expansion factor is
\bea
A \simeq 1.35.
\ena
This value is again too small to explain the cosmological problems.
The conditions~\p{cond21} and \p{cond22} depend only on $t-t_2$,
so changing $t_2$ simply shifts the evolution of the spacetime and
does not give any difference.

\begin{figure}[htb]
\begin{center}
\setlength{\unitlength}{.7mm}
\begin{picture}(50,45)(20,5)
\includegraphics[width=6cm]{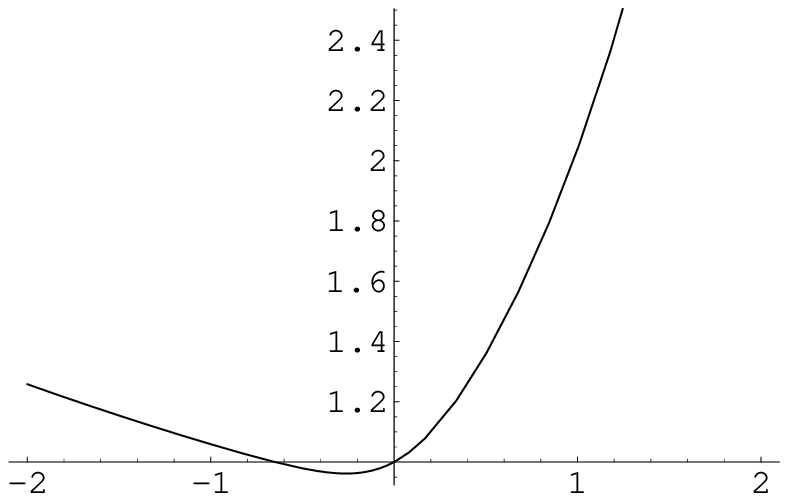}
\put(-38,50){\footnotesize $\d(t)$}
\put(5,3){\footnotesize $t$}
\end{picture}
\caption{The behavior of the size of the flat internal space $\d(t)$
in~\p{mfactor}.}
\label{f9}
\end{center}
\end{figure}

The behavior of the size of the internal space is also shown in Fig.~\ref{f9}.
There is no stable point in the size of the internal space, and in the
infinite future its size goes to infinity.

\subsection{Spherical internal space}

The conditions~\p{cond1} and \p{cond2} for $n=7$ and spherical internal
space $\s=+1$ are (again shifting the time to set $t_1=0$)
\bea
\label{cond31}
&& n_4(t) \equiv \frac{3}{4} \tanh[3c(t-t_2)]
- \frac{\sqrt{21}}{4} \tanh(3\sqrt{3/7}\; ct)>0, \\
&& \frac{9}{8}\left(\frac{1}{\cosh^2[3c(t-t_2)]}
- \frac{1}{\cosh^2(3\sqrt{3/7}\; ct)}\right) - n_4^2(t) > 0.
\label{cond32}
\ena

\begin{figure}[htb]
\begin{minipage}{.45\linewidth}
\begin{center}
\setlength{\unitlength}{.7mm}
\begin{picture}(50,45)(20,5)
\includegraphics[width=6cm]{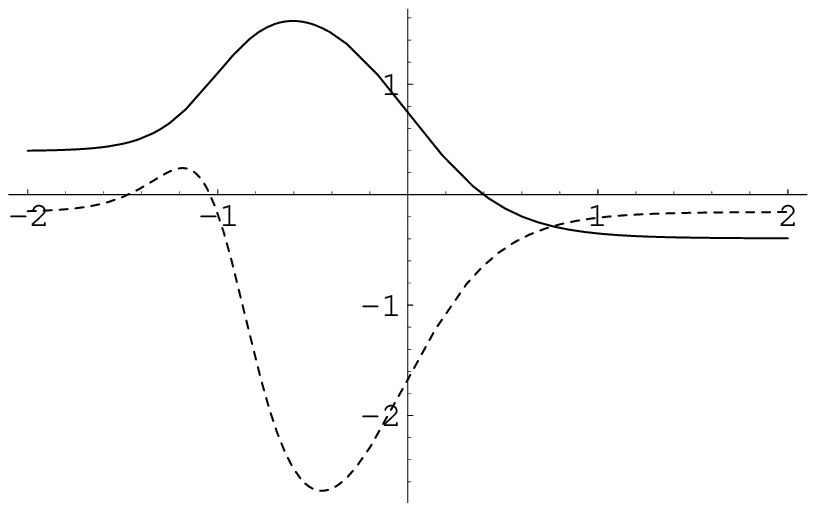}
\put(5,30){\footnotesize $t$}
\end{picture}
\caption{The lhs of eq.~\p{cond31} (solid line) and \p{cond32} (dashed line)
for $t_2=-1$.}
\label{f10}
\end{center}
\end{minipage}
\hspace{3mm}
\begin{minipage}{.5\linewidth}
\begin{center}
\setlength{\unitlength}{.7mm}
\begin{picture}(50,45)(15,5)
\includegraphics[width=6cm]{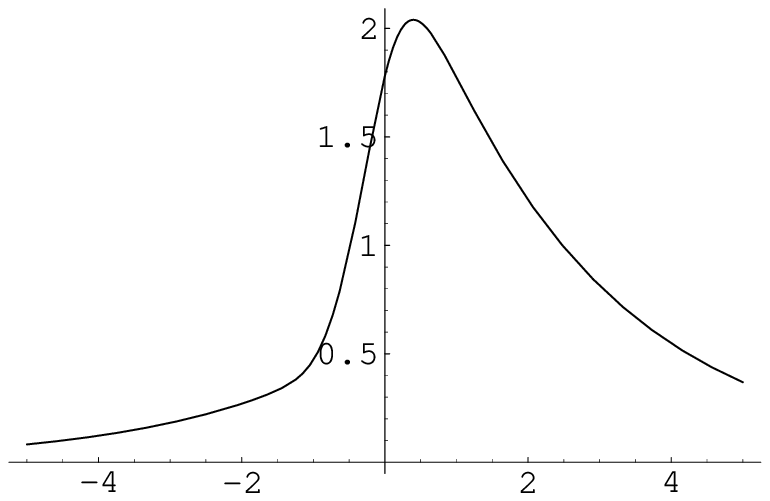}
\put(-30,50){\footnotesize $S(t)$}
\put(5,2){\footnotesize $t$}
\end{picture}
\caption{The behavior of the scale factor $S(t)$ in~\p{mfactor} for $t_2=-1$.}
\label{f11}
\end{center}
\end{minipage}
\end{figure}

The time ranges of $\eta$ and $t$ are the same as the flat internal space.
We have reported the result for $t_2=0$ in ref.~\cite{NO1};
the result indicates that there is no period of accelerated expansion.
We have examined what happens if we change the parameter $t_2$.
The results for $c=1$ and $t_2=-1$ are shown in Fig.~\ref{f10}.
We find that there is a certain period of negative time that the
conditions~\p{cond31} and \p{cond32} are satisfied, though the
universe begin contraction after some positive time. The behavior of
the scale factor $S(t)$ in~\p{mfactor} is depicted in Fig.~\ref{f11}.
It contracts on both ends $t\to \pm \infty$.
The acceleration starting time $T_1$ and ending time $T_2$ are read off from
Fig.~\ref{f10} as
\bea
T_1 \simeq -1.48, \qquad
T_2 \simeq -1.04,
\ena
and the expansion factor is
\bea
A \simeq 1.35.
\ena
This value is again too small to explain the cosmological problems.
We have also checked that the typical behavior for negative $t_2$ is again
basically the same as $t_2=-1$ case, but the period of the accelerated
expansion and the value of the expansion factor during the accelerated
expansion change.

\begin{figure}[htb]
\begin{minipage}{.45\linewidth}
\begin{center}
\setlength{\unitlength}{.7mm}
\begin{picture}(50,45)(20,5)
\includegraphics[width=6cm]{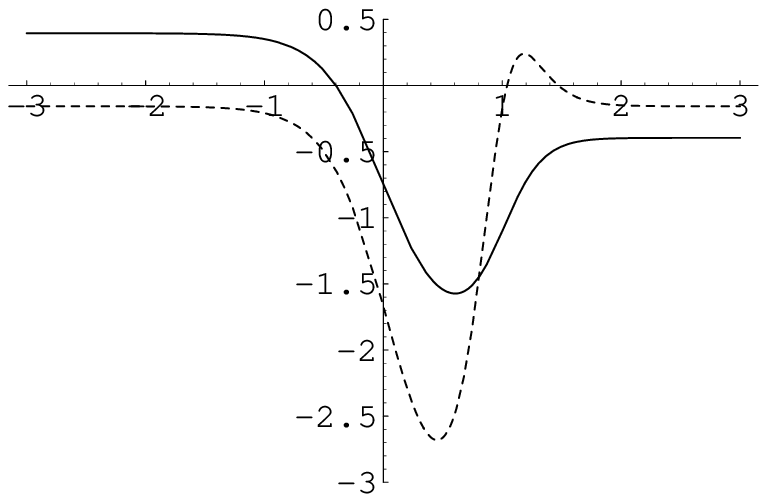}
\put(3,45){\footnotesize $t$}
\end{picture}
\caption{The lhs of eq.~\p{cond31} (solid line) and \p{cond32} (dashed line)
for $t_2=+1$.}
\label{f12}
\end{center}
\end{minipage}
\hspace{3mm}
\begin{minipage}{.5\linewidth}
\begin{center}
\setlength{\unitlength}{.7mm}
\begin{picture}(50,45)(20,5)
\includegraphics[width=6cm]{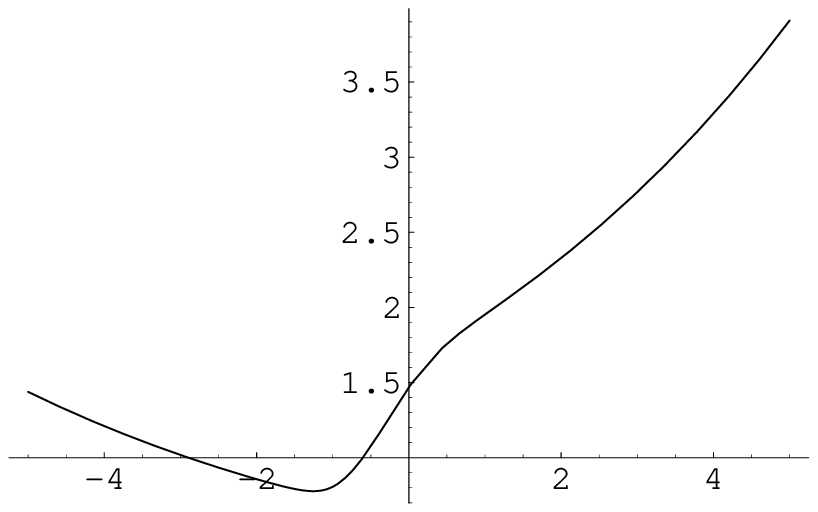}
\put(-40,50){\footnotesize $\d(t)$}
\put(5,5){\footnotesize $t$}
\end{picture}
\caption{The behavior of the size of the flat internal space $\d(t)$
in~\p{mfactor} for $t_2=-1$.}
\label{f13}
\end{center}
\end{minipage}
\end{figure}

A different behavior is observed for positive $t_2$. We find the
acceleration occurs while the universe is already contracting for $t>0$,
as shown in Fig.~\ref{f12}. We do not know if this case gives any
interesting cosmology.
The behavior of the size of the internal space is also shown for $t_2=-1$
in Fig.~\ref{f13}.
There is no stable point in the size of the internal space, and in the
infinite future its size goes to infinity. There is no significant change
in this behavior if we further change the parameter $t_2$.

\section{SD2-brane}

In this section, we discuss accelerating cosmologies from SD2-brane.

\subsection{SD2-brane obtained from SM2-brane}

We first show how the SD2-brane can be obtained by dimensional reduction
from SM2-branes. Though it is possible to examine general dimensions,
we restrict ourselves to $d=11$ here. We will see why the qualitative
behavior is similar to SM2-brane.

The 10-dimensional metric in the string frame is obtained from the
11-dimensional one by the relation
\bea
ds_{11}^2 = e^{-2\phi/3} ds_{10,s}^2 + e^{4\phi/3} dx_{10}^2.
\label{11metric}
\ena
In our solutions~\p{oursol}, we consider SM2-brane by taking $q=2$,
but we set $p=4$ with one extra coordinate outside SM2-brane, in which
direction we make dimensional reduction. We choose $x_1,x_2,x_3$ to be
the SD2-brane world-volume and $x_4$ the direction of dimensional reduction.
We further set $c\equiv c_1=c_2=c_3, c' \equiv c_1'=c_2'=c_3'$ but leave
$c_4$ and $c_4'$ arbitrary. There is no dilaton which comes from the metric
as~\p{11metric} upon dimensional reduction, and we put $a=c_\phi=0$.
This gives
\bea
\phi &=& \frac{1}{4} \ln [\cosh 3c (t-t_2)] + \frac{3}{4} (c+2c_4)t
+ \frac{3}{2} c'_4, \nn
ds_{10,s}^2 &=& e^{2\phi/3} [\cosh 3c (t-t_2)]^{1/3}\Bigg[
e^{-(c+2c_4)t/5 - 2(3c'+c_4')/5} \left\{ - e^{12 g(t)} dt^2
+ e^{2g(t)}d\Sigma_{6,\s}^2\right\} \nn
&& \hs{20} +\; [\cosh 3c(t-t_2)]^{-1} e^{2c'} d{\bf x}^2 \Bigg].
\ena
In the Einstein frame $ds_{10,E}^2 = e^{-\phi/2} ds_{10,s}^2$, this reduces to
\bea
ds_{10,E}^2 &=& [\cosh 3c (t-t_2)]^{3/8}\Bigg[ e^{-3(c+2c_4)t/40-3(8c'
+c_4')/20} \left\{ - e^{12 g(t)} dt^2 + e^{2g(t)}d\Sigma_{6,\s}^2
\right\} \nn
&& \hs{20} +\; [\cosh 3c(t-t_2)]^{-1} e^{(c+2c_4)t/8+(8c'+c_4')/4} d{\bf x}^2
\Bigg].
\ena
The relation to the notation of \cite{CGG} is given by
\bea
c=\frac{16}{15} \a, \quad
c_4=\frac{5}{8} c_1 -\frac{8}{15}\a,
\ena
which must obey \p{condconst} or
\bea
30 \b^2 = \frac{128}{25}\a^2 + \frac{15}{32} c_1^2.
\ena

The scale factor $S(t)$ and $\d(t)$ are given as
\bea
\label{dfactor}
\d(t) &=& [\cosh (16\a/5)(t-t_2)]^{3/16} e^{g(t) -3c_1t/64 - 3\tilde c'/4},\nn
S(t) &=& [\cosh (16\a/5)(t-t_2)]^{1/4} e^{3g(t) -c_1 t/16 - \tilde c'},\\
\phi &=& \frac{1}{4}\ln[\cosh(16\a/5)(t-t_2)] +\frac{15}{16}c_1 t +c_\phi',
\nonumber
\ena
where we have also put
\bea
\tilde c'= \frac{8c'+c_4'}{10}, \quad
c_\phi' = \frac{3}{2} c_4'.
\ena
On the other hand, if we consider the SD2-brane with $d=10, q=2, p=3, n=6,
a=\frac{1}{2},\e=+1$ in eq.~\p{oursol}, we get precisely \p{dfactor}
with different parametrization
\bea
c=\a + \frac{5}{64}c_1, \quad
c_\phi = \frac{15}{16}c_1 - \frac{4}{5}\a, \quad
c' = \frac{5}{4} \tilde c'.
\ena
The scale factors~\p{dfactor} almost coincide with those for SM2-brane
in~\p{mfactor} if we choose $n=6$, though the power of the $\cosh$ is
slightly different and there is an additional $c_1 t$ term in the
exponent. However, the difference is small and one may expect
that qualitative features remain the same. This is the reason why one finds
basically the same behavior as SM2-brane. To confirm this, we present
some results for SD2-brane below.

Continuing this reduction, one should leave the constants $c_\a$ arbitrary
in order to get general lower-dimensional solutions;
in the SM2-brane solution of ref.~\cite{CGG}, homogeneity of the space
outside the S-brane is assumed, and then general solutions cannot be
obtained by dimensional reduction. If we start with our solutions without
putting any relations among $c_\a$'s, we can simultaneously discuss
all possible cases. This is another virtue of our solutions, in addition
to the fact that they cover the vacuum solution.

\subsection{Hyperbolic internal space}

The conditions~\p{cond1} and \p{cond2} give
\bea
\label{cond41}
&& n_5(t) \equiv \frac{4}{5}\a \tanh[(16\a/5)(t-t_2)] - 3\b \coth(5\b t)
-\frac{c_1}{16} >0, \\
&& \frac{32\a^2/25}{\cosh^2[(16\a/5)(t-t_2)]}
+ \frac{15\b^2/2}{\sinh^2(5\b t)} - n_5^2(t) > 0,
\label{cond42}
\ena
where we have again set $t_1=0$ by a shift of time.

\begin{figure}[htb]
\begin{minipage}{.45\linewidth}
\begin{center}
\setlength{\unitlength}{.7mm}
\begin{picture}(50,45)(20,5)
\includegraphics[width=6cm]{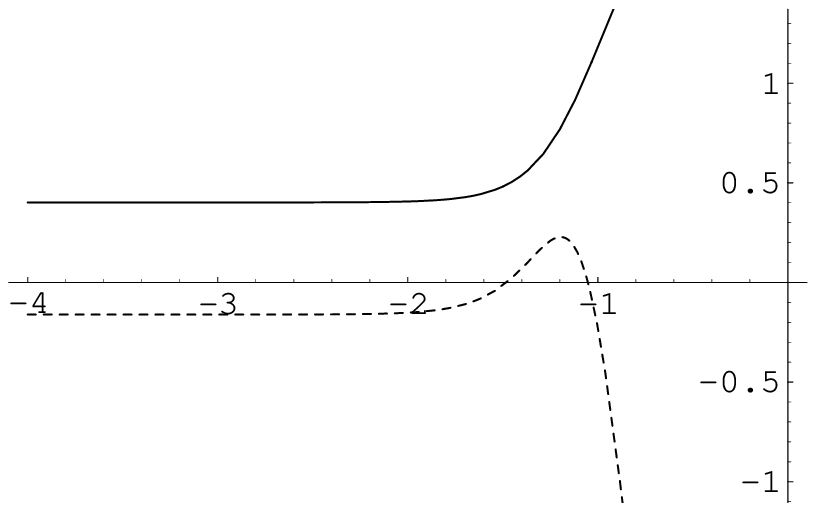}
\put(5,22){\footnotesize $t$}
\end{picture}
\caption{The lhs of eq.~\p{cond41} (solid line) and \p{cond42} (dashed line)
for $t_2=-1$.}
\label{f14}
\end{center}
\end{minipage}
\hspace{3mm}
\begin{minipage}{.5\linewidth}
\begin{center}
\setlength{\unitlength}{.7mm}
\begin{picture}(50,45)(15,5)
\includegraphics[width=6cm]{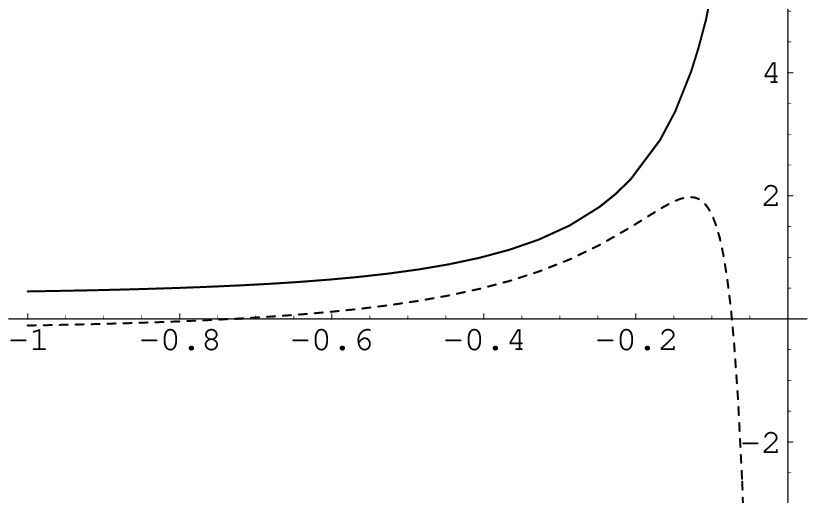}
\put(5,20){\footnotesize $t$}
\end{picture}
\caption{The lhs of eq.~\p{cond41} (solid line) and \p{cond42} (dashed line)
for $t_2=+1$.}
\label{f15}
\end{center}
\end{minipage}
\end{figure}

The conditions~\p{cond41} and \p{cond42} for $\a=0.92, c_1=1,\b=\frac{2}{5}$
was examined in \cite{R}, and the qualitative behavior is found to be the
same as SM2-brane, as expected. We have studied how the behavior changes
if we change parameter $t_2$. The results for $t_2=-1,+1$ are shown
in Figs.~\ref{f14} and \ref{f15}, respectively.
The expansion factor is also obtained for $t_2=+1$ as
\bea
A \simeq 3.28.
\ena
Again similarly to the SM2-brane, the positive $t_2$ gives bigger $A$,
but not so much change. We thus see that the behavior does not change
qualitatively from the case of SM2-brane, as expected.

However, there is a potential subtlety in this picture of SD2-brane.
This is because the string coupling is given by the dilaton as $e^\phi$,
and it may diverge for the solution~\p{dfactor} for certain range of
parameters. If this happens, quantum effects dominate and we cannot simply
trust the result. Fortunately there is a safe range of parameters for this
compactification on hyperbolic internal space because the time range is
restricted to $t<0$. From eq.~\p{dfactor}, one finds that if the condition
\bea
c_1 >\frac{64}{75}|\a|
\ena
is satisfied, we can remain in the weak coupling region and the above picture
may be trusted. The above choice of the parameters is within this range
so the above result may be correct. However we will see that this may
be a more serious problem for other internal spaces.

\subsection{Flat internal space}

The conditions~\p{cond1} and \p{cond2} give
\bea
\label{cond43}
&& n_6(t) \equiv \frac{4}{5}\a \tanh[(16\a/5)(t-t_2)] + 3\b -\frac{c_1}{16}
>0, \\
&& \frac{32\a^2/25}{\cosh^2[(16\a/5)(t-t_2)]} - n_6^2(t) > 0.
\label{cond44}
\ena
This case for $\a=0.92, c_1=1,\b=\frac{2}{5}$ was examined in \cite{R}
with results similar to SM2-brane again, and one can see that the change
of $t_2$ simply gives the shift in time of the whole behavior without
new feature.

However, in this case the time range is from $-\infty$ to $+\infty$ so that
the dilaton~\p{dfactor} always gives strong string coupling somewhere for
large $|t|$ whatever the choice of the parameters are, and it is not clear
if this classical analysis is valid.
So we cannot conclude whether this model gives a reasonable result.
It is actually known that this limit is the one in which 11-th dimension
becomes large~\cite{WI}. This case is better described in the SM2-brane
picture, with large 11-th dimension. This is what we have already analyzed
in the preceding section.

\subsection{Spherical internal space}

The conditions~\p{cond1} and \p{cond2} give
\bea
\label{cond45}
&& n_7(t) \equiv \frac{4}{5}\a \tanh[(16\a/5)(t-t_2)] - 3\b\tanh(5\b t)
-\frac{c_1}{16} > 0, \\
&& \frac{32\a^2/25}{\cosh^2[(16\a/5)(t-t_2)]}-\frac{15\b^2/2}{\cosh^2(5\b t)}
- n_7^2(t) > 0,
\label{cond46}
\ena
where we have again set $t_1=0$ by a shift of time.

\begin{figure}[htb]
\begin{minipage}{.45\linewidth}
\begin{center}
\setlength{\unitlength}{.7mm}
\begin{picture}(50,45)(20,5)
\includegraphics[width=6cm]{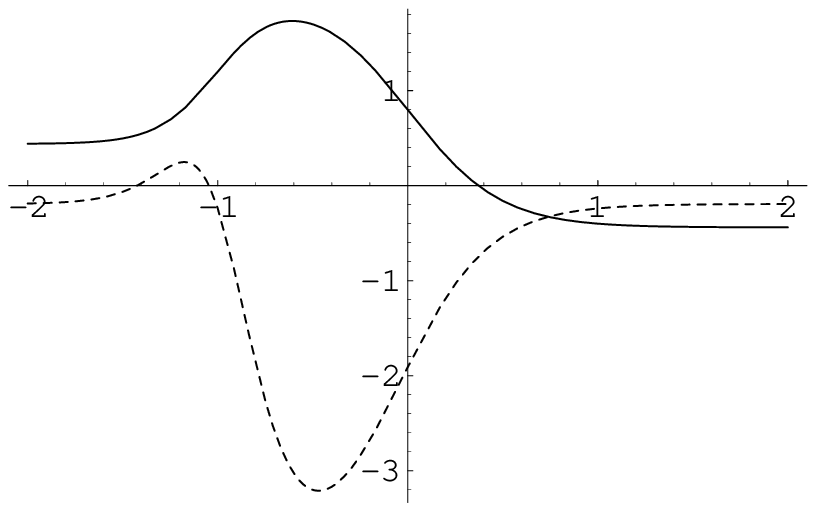}
\put(5,32){\footnotesize $t$}
\end{picture}
\caption{The lhs of eq.~\p{cond45} (solid line) and \p{cond46} (dashed line)
for $t_2=-1$.}
\label{f16}
\end{center}
\end{minipage}
\hspace{3mm}
\begin{minipage}{.5\linewidth}
\begin{center}
\setlength{\unitlength}{.7mm}
\begin{picture}(50,45)(15,5)
\includegraphics[width=6cm]{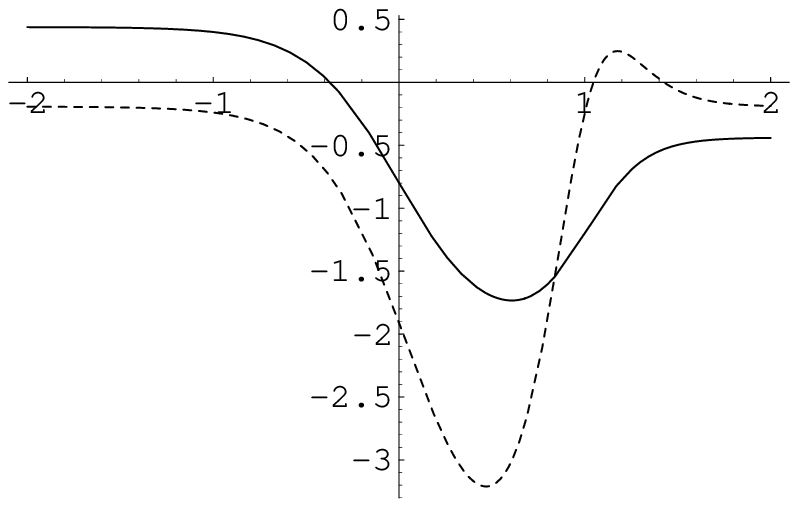}
\put(5,44){\footnotesize $t$}
\end{picture}
\caption{The lhs of eq.~\p{cond45} (solid line) and \p{cond46} (dashed line)
for $t_2=+1$.}
\label{f17}
\end{center}
\end{minipage}
\end{figure}

We have examined the behaviors for $\a=1, c_1=0,\b=\frac{8}{5\sqrt{15}}$
for $t_2=-1,+1$ which are shown in Figs.~\ref{f16} and \ref{f17}, respectively.
We find again that there is a possibility of getting accelerating cosmologies
for negative $t_2$, and also for positive $t_2$ though in this case the
universe is contracting. The typical features of the solution is essentially
the same as SM2-brane case. Here again this internal space suffers from
the problem of strong string coupling.
It is not clear if we can trust this result, but the analysis in SM2-brane
seems to support the basic behavior.

\section{Conclusions and discussions}

We have examined accelerating cosmologies obtained from time-dependent
solutions in superstring/M theories including parameter dependence,
with the hope to get bigger expansion factors.
As a new feature, we have found that it is also possible to obtain
accelerating phase not only for hyperbolic and flat internal spaces but
also for spherical space. In the last case, the universe eventually enters
into the contracting phase.

The obtained expansion factors of our universe seem to be rather small to
solve the horizon and flatness problems, and this seems to be typical
problem in these models. This factor improves slightly if we change
parameters of the models, but still not enough.

We have also shown the relation between SM2-brane and SD2-brane,
and thus have given the reason why the qualitative behavior is
basically the same as SM2-brane. There is also a problem associated with
the strong string coupling in SD2-brane, in which case one must return
to the SM2-brane picture.

One of the common features of the above results is that our spacetime starts
from very small size while the internal space is fairly large, and our
scale factor grows in time while the size of the internal space shrinks
for a while (perhaps in our time). It is interesting to further pursue
this possibility.

An obvious generalization of our work would be to consider more general
case allowing the full parameters in our solution~\p{oursol}.
Also it would be interesting to examine other possibility of the internal
space such as product space. It is extremely important to investigate how
one can achieve expansion factor large enough to explain the cosmological
problems in this context.

\section*{Acknowledgements}

This work was supported in part by Grants-in-Aid for Scientific Research
Nos. 12640270 and 02041.

\newcommand{\NP}[1]{Nucl.\ Phys.\ B\ {\bf #1}}
\newcommand{\PL}[1]{Phys.\ Lett.\ B\ {\bf #1}}
\newcommand{\CQG}[1]{Class.\ Quant.\ Grav.\ {\bf #1}}
\newcommand{\CMP}[1]{Comm.\ Math.\ Phys.\ {\bf #1}}
\newcommand{\IJMP}[1]{Int.\ Jour.\ Mod.\ Phys.\ {\bf #1}}
\newcommand{\JHEP}[1]{JHEP\ {\bf #1}}
\newcommand{\PR}[1]{Phys.\ Rev.\ D\ {\bf #1}}
\newcommand{\PRL}[1]{Phys.\ Rev.\ Lett.\ {\bf #1}}
\newcommand{\PRE}[1]{Phys.\ Rep.\ {\bf #1}}
\newcommand{\PTP}[1]{Prog.\ Theor.\ Phys.\ {\bf #1}}
\newcommand{\PTPS}[1]{Prog.\ Theor.\ Phys.\ Suppl.\ {\bf #1}}
\newcommand{\MPL}[1]{Mod.\ Phys.\ Lett.\ {\bf #1}}
\newcommand{\JP}[1]{Jour.\ Phys.\ {\bf #1}}


\end{document}